\documentstyle[11pt]{article}

\makeatletter

\let\@eqnsel\hfil
\newskip\@eqnmar
\@eqnmar\@centering
\let\@eqnhook\relax
\def\eqnarray{\stepcounter{equation}\let\@currentlabel=\theequation
\global\@eqnswtrue
\global\@eqcnt\z@\tabskip\@eqnmar\let\\=\@eqncr
\@eqnhook
$$\halign to \displaywidth\bgroup\@eqnsel
  $\displaystyle\tabskip\z@{##}$&\global\@eqcnt\@ne
  \hfil${{}##{}}$\hfil&\global\@eqcnt\tw@
  $\displaystyle{##}$\hfil
  \tabskip\@centering&\llap{##}\tabskip\z@\cr}
\def\lefteqn#1{\hbox to 4\arraycolsep{$\displaystyle #1$\hss}}
\@ifundefined{mathindent}{}{\@eqnmar\mathindent
    \def\@eqnhook{\abovedisplayskip\topsep
    \ifvmode\advance\abovedisplayskip\partopsep\fi
    \belowdisplayskip\abovedisplayskip
    \belowdisplayshortskip\abovedisplayskip
    \abovedisplayshortskip\abovedisplayskip}}

\def\[{\relax\ifmmode\@badmath\else
 \begin{trivlist}
 \@beginparpenalty\predisplaypenalty
 \@endparpenalty\postdisplaypenalty
 \item[]\leavevmode
 \hbox to\linewidth\bgroup$ \displaystyle
 \hskip\mathindent\bgroup\fi}
\def\]{\relax\ifmmode \egroup $\hfil \egroup \end{trivlist}\else \@badmath \fi}
\def\equation{\@beginparpenalty\predisplaypenalty
 \@endparpenalty\postdisplaypenalty
\refstepcounter{equation}\trivlist \item[]\leavevmode
 \hbox to\linewidth\bgroup $ \displaystyle
\hskip\mathindent}
\def\endequation{$\hfil \displaywidth\linewidth\@eqnnum\egroup \endtrivlist}
\def\eqnarray{\stepcounter{equation}\let\@currentlabel=\theequation
\global\@eqnswtrue
\global\@eqcnt\z@\tabskip\mathindent\let\\=\@eqncr
\abovedisplayskip\topsep\ifvmode\advance\abovedisplayskip\partopsep\fi
\belowdisplayskip\abovedisplayskip
\belowdisplayshortskip\abovedisplayskip
\abovedisplayshortskip\abovedisplayskip
$$\halign to
\linewidth\bgroup\@eqnsel\hskip\@centering$\displaystyle\tabskip\z@
 {##}$&\global\@eqcnt\@ne \hskip 2\arraycolsep \hfil${##}$\hfil
 &\global\@eqcnt\tw@ \hskip 2\arraycolsep $\displaystyle{##}$\hfil
 \tabskip\@centering&\llap{##}\tabskip\z@\cr}
\def\endeqnarray{\@@eqncr\egroup
 \global\advance\c@equation\m@ne$$\global\@ignoretrue }
\newdimen\mathindent
\mathindent = \leftmargini

\newtoks\@stequation
\def\subequations{\refstepcounter{equation}%
  \edef\@savedequation{\the\c@equation}%
  \@stequation=\expandafter{\theequation}
  \edef\@savedtheequation{\the\@stequation}
  \edef\oldtheequation{\theequation}%
  \setcounter{equation}{0}%
  \def\theequation{\oldtheequation\alph{equation}}}
\def\endsubequations{%
  \ifnum\c@equation < 2 \@warning{Only \the\c@equation\space subequation
    used in equation \@savedequation}\fi%
  \setcounter{equation}{\@savedequation}%
  \@stequation=\expandafter{\@savedtheequation}%
  \edef\theequation{\the\@stequation}%
  \global\@ignoretrue}

\newcommand{\p}{\partial}
\newcommand{\del}{\nabla}
\newcommand{\Lag}{{\cal L}}
\newcommand{\ED}{{\cal E}}
\newcommand{\ess}{{\cal S}}
\newcommand{\tends}{\rightarrow}
\newcommand{\E}{{\rm e}}
\newcommand{\I}{{\rm i}}
\newcommand{\R}{\V{R}}
\newcommand{\Z}{\V{Z}}
\newcommand{\hsp}[1]{\hspace{#1mm}}
\newcommand{\msp}[1]{\mbox{\hsp{#1}}}
\newcommand{\vsp}[1]{\vspace{#1mm}}
\newcommand{\ts}{\textstyle}
\newcommand{\ds}{\displaystyle}
\newcommand{\dee}{\mbox{d}}
\newcommand{\diff}[2]{{\ds\frac{\dee #1}{\dee #2}}}
\newcommand{\pdiff}[2]{{\ds\frac{\p #1}{\p #2}}}
\newcommand{\arrow}{\rightarrow}
\newcommand{\V}[1]{{\bf #1}}
\newcommand{\tw}[1]{\tilde{#1}}
\newcommand{\dn}{\mbox{dn}}
\newcommand{\sn}{\mbox{sn}}
\newcommand{\cn}{\mbox{cn}}
\renewcommand{\Re}{\mbox{Re}\,}
\renewcommand{\Im}{\mbox{Im}\,}
\newcommand{\recip}[1]{\frac{1}{#1}}
\newcommand{\half}{\recip{2}}
\newcommand{\quar}{\recip{4}}
\newcommand{\II}{I\hsp{-1}I}
\newenvironment{Vector}{\left(\!\begin{array}{c}}{\end{array}\!\right)}
\newcommand{\bve}{\begin{Vector}}
\newcommand{\eve}{\end{Vector}}

\newcommand{\fn}[1]{\footnote[2]{#1}\hsp{-3}}
\newcommand{\beq}[1]{\begin{equation}\label{#1}}
\newcommand{\eeq}{\end{equation}}
\newcommand{\bsub}[1]{\begin{subequations}\label{#1}}
\newcommand{\esub}{\end{subequations}}
\newcommand{\bea}{\begin{eqnarray}}
\newcommand{\eea}{\end{eqnarray}}
\newcommand{\eqnum}[1]{(\ref{#1})}
\newcommand{\Sec}[1]{Section~\ref{#1}}
\newcommand{\ie}{$i.\:e.$~}
\newcommand{\comma}{\,,\msp{6}}

\def\citen#1{%
\if@filesw \immediate \write \@auxout {\string \citation {#1}}\fi
\@tempcntb\m@ne \let\@h@ld\relax \def\@citea{}%
\@for \@citeb:=#1\do {%
  \@ifundefined {b@\@citeb}%
    {\@h@ld\@citea\@tempcntb\m@ne{\bf ?}%
    \@warning {Citation `\@citeb ' on page \thepage \space undefined}}%
    {\@tempcnta\@tempcntb \advance\@tempcnta\@ne
    \setbox\z@\hbox\bgroup
    \ifnum0<0\csname b@\@citeb \endcsname \relax
	\egroup \@tempcntb\number\csname b@\@citeb \endcsname \relax
	\else \egroup \@tempcntb\m@ne \fi
    \ifnum\@tempcnta=\@tempcntb
	\ifx\@h@ld\relax
	  \edef \@h@ld{\@citea\csname b@\@citeb\endcsname}%
	\else
	  \edef\@h@ld{\hbox{--}\penalty\@highpenalty
            \csname b@\@citeb\endcsname}%
	\fi
    \else
	\@h@ld\@citea\csname b@\@citeb \endcsname
	\let\@h@ld\relax
    \fi}%
 \def\@citea{,\penalty\@highpenalty\hskip.13em plus.1em minus.1em}%
}\@h@ld}
\def\cite{\leavevmode\unskip\@ifnextchar[{\@tempswatrue\@citew}%
            {\@tempswafalse\@citex}}
\def\@citew[#1]#2{\ifnum\lastpenalty=\z@ \penalty\@highpenalty \fi
   \ [{\multiply\@highpenalty 3
   \citen{#2}},\penalty\@highpenalty\ #1]\spacefactor\@m}
\def\@citex#1{\begingroup\leavevmode\unskip
  \def\@tempa{\@cite{\citen{#1}}\endgroup}\futurelet\@tempb\@citey}%
\def\@citey{\let\@tempc\@tempa
   \ifx\@tempb.\ifnum\spacefactor>2999 \let\@tempb\relax\fi\let\@tempc\@citez
   \else\ifx\@tempb,\let\@tempc\@citez
   \else\ifx\@tempb:\let\@tempc\@citez
   \else\ifx\@tempb;\let\@tempc\@citez
   \fi\fi\fi\fi
   \@tempc}%
\def\@citez#1{\@tempb\futurelet\@tempb\@citey}%
\def\@cite#1{$\m@th\the\scriptfont\z@\edef\bf{\the\scriptfont\bffam}%
	^{\hbox{#1}}$}

\def\baselinestretch{2}
\def\setstretch#1{\renewcommand{\baselinestretch}{#1}}
\def\@setsize#1#2#3#4{\@nomath#1%
   \let\@currsize#1\baselineskip
   #2\baselineskip\baselinestretch\baselineskip
   \parskip\baselinestretch\parskip
   \setbox\strutbox\hbox{\vrule height.7\baselineskip
	depth.3\baselineskip width\z@}%
   \normalbaselineskip\baselineskip#3#4}
\skip\footins 20pt plus4pt minus4pt
\def\@xfloat#1[#2]{\ifhmode \@bsphack\@floatpenalty -\@Mii\else
   \@floatpenalty-\@Miii\fi\def\@captype{#1}\ifinner
	\@parmoderr\@floatpenalty\z@
    \else\@next\@currbox\@freelist{\@tempcnta\csname ftype@#1\endcsname
	\multiply\@tempcnta\@xxxii\advance\@tempcnta\sixt@@n
	\@tfor \@tempa :=#2\do
			{\if\@tempa h\advance\@tempcnta \@ne\fi
			 \if\@tempa t\advance\@tempcnta \tw@\fi
			 \if\@tempa b\advance\@tempcnta 4\relax\fi
			 \if\@tempa p\advance\@tempcnta 8\relax\fi
	 }\global\count\@currbox\@tempcnta}\@fltovf\fi
    \global\setbox\@currbox\vbox\bgroup
    \def\baselinestretch{1}\small\normalsize
    \boxmaxdepth\z@
    \hsize\columnwidth \@parboxrestore}
\long\def\@footnotetext#1{\insert\footins{\def\baselinestretch{1}\footnotesize
    \interlinepenalty\interfootnotelinepenalty
    \splittopskip\footnotesep
    \splitmaxdepth \dp\strutbox \floatingpenalty \@MM
    \hsize\columnwidth \@parboxrestore
   \edef\@currentlabel{\csname p@footnote\endcsname\@thefnmark}\@makefntext
    {\rule{\z@}{\footnotesep}\ignorespaces
	#1\strut}}}

\@addtoreset{equation}{section}
\def\theequation{\thesection.\arabic{equation}}

\makeatother

\begin{document}
\title{{\Large\sc Time-Independent Solutions to the Two-Dimensional Non-Linear
O(3) Sigma
Model and Surfaces of Constant Mean Curvature} \\ \vsp{3}
{\large{\bf Short title: } CMC Surfaces and the O(3) Model}}
\author{Michael S. Ody\thanks{Email: mso@ukc.ac.uk} \,\,and
	Lewis H. Ryder\thanks{Email: lhr@ukc.ac.uk} \\ \\
	The Physics Laboratory, University of Kent at Canterbury, \\
	Canterbury CT2 7NR, England.}
\date{\today}
\maketitle

\begin{abstract}
It is shown that time-independent solutions to the (2+1)-dimensional non-linear
O(3) sigma
model may be placed in correspondence with surfaces of constant mean curvature
in three-%
dimensional Euclidean space.  The tools required to establish this
correspondence are
provided by the classical differential geometry of surfaces.  A
constant-mean-curvature
surface induces a solution to the O(3) model through the identification of the
Gauss map, or
normal vector, of the surface with the field vector of the sigma model.  Some
explicit
solutions, including the solitons and antisolitons discovered by Belavin and
Polyakov, and a
more general solution due to Purkait and Ray, are considered and the surfaces
giving rise to
them are found explicitly.  It is seen, for example, that the Belavin-Polyakov
solutions are
induced by the Gauss maps of surfaces which are conformal to their spherical
images, \ie
spheres and minimal surfaces, and that the Purkait-Ray solution corresponds to
the family of
constant-mean-curvature helicoids first studied by do Carmo and Dajczer in
1982.  A
generalisation of this method to include time-dependence may shed new light on
the r\^{o}le
of the Hopf invariant in this model.
\end{abstract}

\newpage
\section{Introduction}
\label{intro}

It has been known for a long time that there is a close link between certain
non-linear
partial differential equations, or `evolution equations', and the classical
differential
geometry of surfaces in three-dimensional Euclidean space.  These non-linear
equations arise
from `compatibility conditions' imposed on the {\it linear} Gauss-Weingarten
equations which
describe the surface\cite{eisenhart}.  For example, the sine-Gordon equation is
naturally
associated with pseudospherical surfaces, and the Liouville equation with
spherical and
minimal surfaces.  This situation, where a non-linear compatibility condition
follows from an
over-determined linear system, is exactly mirrored in the `inverse scattering
method' due to
Gardner {\it et al}\cite{gardneretal67}.  It was shown by Lund\cite{lund78},
building on
earlier results \cite{lundregge76}, that this method may be combined with
classical surface
theory to give a geometric interpretation to the procedure of finding the
linear system of
which a given evolution equation is the compatibility condition.  A general
class of such
equations\,---\,the `AKNS system' \cite{ablowitzetal74}, which contains the
sine-Gordon,
Korteweg-de Vries (KdV), modified KdV, and non-linear Schr\"{o}dinger equations
as special
cases\,---\,has been considered in this geometric light
\cite{gursesnutku81,sym82,sym85}.

In this paper, we apply the tools of classical surface theory to a popular
non-linear sigma
model.  The O(3) model in 2+1 dimensions has received much attention as a
result of its
resemblance to (3+1)-dimensional non-Abelian gauge theories (see, for example,
refs.\
\citen{balachandranetal} and \citen{zak}).  The approach adopted here is
somewhat different
from that used by Lund\cite{lund78} in that we regard the (non-linear) field
equation of the
O(3) model not as the compatibility condition of a surface but rather as a
constraint on the
normal vector to the surface.  The normal, or, more correctly, the Gauss map
(described in
\Sec{geom}), is identified with the field vector of the O(3) model.  In this
way a
correspondence is defined between surfaces in three dimensions and solutions to
the field
equation of the model.  Similar techniques have been applied in the past
to the non-linear O(4) model in two dimensions\cite{lund77} and to the O(3)
model
in 1+1 dimensions\cite{houetal85}.  However, there are differences between the
approaches
used in these papers
and the work presented here.  In ref.\ \citen{lund77} the field vector of the
O(4)
model was identified with the {\it position} vector, rather than the normal
vector, of a surface
and the problem led back to the sine-Gordon case of the AKNS system.  In ref.\
\citen{houetal85}
the field vector was equated with the surface normal, as is done here, but a
parametrisation of the surface\,---\,the so-called `asymptotic'
coordinates\,---\,was adopted
which
obscured the variety and interrelationships of the model's solutions.  Namely,
it was found
that the surfaces to which the field vector is normal were all pseudospheres.
By contrast,
it will be shown below that the use of {\it isothermal} coordinates leads to a
much more
faithful correspondence between the solutions to the O(3) model and surfaces in
$\R^3$.  An
expanded account of this work is given by Ody\cite{ody93}.

We begin in \Sec{model} by summarising the salient features of the O(3) model
and a number of
its time-independent solutions.  The relevant differential geometry is given in
\Sec{geom}
and then combined with the mathematics of the O(3) model in \Sec{main}.  In
addition to
finding the surfaces associated with particular solutions, it is shown that
quantities in the
sigma model, such as topological charge, energy density, and total energy, can
be directly
identified with surface-geometric quantities.  \Sec{conc} contains a summary
and concluding
remarks.

\newpage
\section{The non-linear O(3) model in two dimensions}
\label{model}

Here we briefly describe the time-independent O(3) sigma model and a number of
solutions to
its field equation.  For further details, see refs.\ \citen{zak} or
\citen{rajaraman}.

The Lagrangian of the model is simply
\beq{10}
\Lag = {\ts\half}\p_i n_a\,\p_i n_a \equiv {\ts\half}\V{n}_{,i}\cdot\V{n}_{,i}
\eeq
where there is a summation over both the internal index $a = 1,2,3$ and the
space
index $i = 1,2$, and the notation $\V{n}_{,i}$ is shorthand for $\p_i\V{n}$.
The field
vector $\V{n}(x_1,x_2) = (n_1,n_2,n_3)$ of the model is a {\it unit} vector in
internal
space as a result of the non-linear constraint
\beq{20}
\V{n}\cdot\V{n} \equiv n_1^2 + n_2^2 + n_3^2 = 1 \mbox{\hsp{4}for all
$x_1,x_2$.}
\eeq
It is therefore sensible to adopt a polar coordinate system in which
\beq{30}
\V{n} = \bve \sin\Theta\cos\Phi \\ \sin\Theta\sin\Phi \\ \cos\Theta \eve
\eeq
where $\Theta$ and $\Phi$ are functions of $x_1$ and $x_2$.  The field manifold
is clearly
the unit sphere $\ess^2$ centred on the origin of internal space $\R^3$.  It
follows from
\eqnum{10} and \eqnum{20} that the Euler-Lagrange equation, or field equation,
of the model
is
\beq{40}
\del^2\V{n} - (\V{n} \cdot \del^2\V{n})\V{n} = 0.
\eeq
Some solutions to this non-linear equation are described below.

It proves convenient to map the field manifold $\ess^2$ onto a plane via a
stereographic
projection from the `south pole' $\V{n} = (0,0,-1)$.  We define
\beq{50}
W_1 \equiv \frac{n_1}{1+n_3} = \tan{\ts\half}\Theta\cos\Phi \comma
W_2 \equiv \frac{n_2}{1+n_3} = \tan{\ts\half}\Theta\sin\Phi
\eeq
where $(W_1,W_2)$ are Cartesian coordinates in the $n_1 n_2$-plane of internal
space.  We can
define a complex function $W(z,z^*)$ by
\beq{60}
W \equiv W_1 + \I W_2 = \frac{n_1 + \I n_2}{1+n_3} =
\tan{\ts\half}\Theta\,\E^{\I\Phi}
\eeq
where $z \equiv x_1 + \I x_2 = r\E^{\I\phi}$.  The inverse projection is given
by
\beq{70}
\V{n} = \frac{1}{1 + |W|^2}\bve 2\,\Re W \\ 2\,\Im W \\ 1 - |W|^2 \eve
\eeq
or, equivalently, by substituting
\beq{80}
\Theta = 2\tan^{-1}|W| \comma \Phi = \arg W
\eeq
into \eqnum{30}.  In terms of $W$, the field equation \eqnum{40} becomes
\beq{90}
\p\p^* W - \frac{2W^*\,\p W\,\p^* W}{1 + |W|^2} = 0
\eeq
where $\p$ and $\p^*$ are shorthand for $\ds\pdiff{}{z}$ and $\ds\pdiff{}{z^*}$
respectively.
The function $W$, instead of $\V{n}$, may be regarded as the fundamental object
in the O(3)
model.

The total energy $E_{\rm tot}$ of a time-independent solution $\V{n}(x_1,x_2)$
is given by
\beq{100}
E_{\rm tot} = \int_{\R^2}\ED\,\dee x_1 \wedge \dee x_2
\eeq
where the energy density $\ED(x_1,x_2)$ is
\beq{110}
\ED = {\ts\half}\V{n}_{,i} \cdot \V{n}_{,i} = -{\ts\half}\V{n} \cdot
\del^2\V{n}.
\eeq
The integral in \eqnum{100} will be finite if $\V{n}$ tends to the same
constant value
$\V{n}_0$ at large distances from the origin of the $x_1 x_2$-plane, \ie if
\beq{120}
\lim_{r\tends\infty}\V{n} = \V{n}_0 \mbox{\hsp{4}for all $\phi$.}
\eeq
This behaviour of the field at infinity endows the coordinate space $\R^2$ with
the topology
of the sphere $\ess^2$.  A field $\V{n}(x_1,x_2)$ which obeys \eqnum{120} is
then a mapping
from this `compactified' coordinate space $\ess^2$ to the field manifold
$\ess^2$.  Such
maps fall into homotopy classes as a result of the homotopy
\beq{130}
\pi_2(\ess^2) \sim \Z.
\eeq
The degree $Q$ of the map $\V{n} : \ess^2 \arrow \ess^2$ is an integer,
positive or negative,
which is the same for all maps in the same homotopy class and is interpreted
physically as a
topological charge.  An expression for $Q$ can be found\cite{flanders} by
integrating the pullback
$\V{n}^*\beta$ of the normalised area element $\beta$ on the field manifold
$\ess^2$:
\beq{140}
\beta = \frac{1}{4\pi}\star\V{n} \cdot \dee\V{n}
      = \frac{1}{8\pi}\,\V{n} \cdot \dee\V{n} \times \dee\V{n}.
\eeq
(Here, $\star$ is the Hodge `star' operator.)  If $\V{n}$ obeys the boundary
condition
\eqnum{120} then we can write
\beq{150}
Q = \int_{\R^2}\V{n}^*\beta \equiv \int_{\R^2}J\,\dee x_1 \wedge \dee x_2
\eeq
where $J(x_1,x_2)$ is the charge density given by
\beq{160}
J = \frac{1}{8\pi}\,\epsilon_{ij}\,\V{n} \cdot \V{n}_{,i} \times \V{n}_{,j}
  = \frac{1}{4\pi}\,\V{n} \cdot \V{n}_{,1} \times \V{n}_{,2}.
\eeq
In terms of the complex function $W$, we have
\beq{170}
\ED = \frac{4(|\p W|^2 + |\p^*W|^2)}{(1 + |W|^2)^2}
\eeq
\beq{180}
J = \frac{|\p W|^2 - |\p^*W|^2}{\pi(1 + |W|^2)^2}.
\eeq
The homotopy classification \eqnum{130} applies only to finite-energy
solutions, \ie those
which obey \eqnum{120}.  It was shown by Belavin and Polyakov\cite{belapoly75}
that such solutions satisfy, in
addition to the field equation \eqnum{40}, the so-called `duality equations'
\beq{190}
\V{n}_{,i} = \pm\,\epsilon_{ij}\,\V{n}\times\V{n}_{,j}.
\eeq
The two equations ($i=1,2$) contained in \eqnum{190} are not independent.  We
will refer to
the $i=1$ case as `the' duality equation:
\beq{200}
\V{n}_{,1} = \pm\,\V{n} \times \V{n}_{,2}.
\eeq
The total energy $E_{\rm tot}$ of a field $\V{n}$ which satisfies \eqnum{200}
is related to
its topological charge $Q$ by
\beq{210}
E_{\rm tot} = 4\pi |Q|.
\eeq
In terms of $W$, the duality equation becomes the Cauchy-Riemann conditions
\bsub{220}
\beq{220a}
W_{1,1} = \mp\,W_{2,2}
\eeq
\beq{220b}
W_{2,1} = \pm\,W_{1,2}
\eeq
\esub
\ie
\beq{230}
W = \omega
\eeq
where $\omega$ is a complex analytic function of either $z$ (lower signs) or
$z^*$ (upper
signs).  This solution \eqnum{230} is the Belavin-Polyakov (BP) solution.  It
was shown by
Garber {\it et al.}\cite{garberetal79}\ that this is the {\it only}
finite-energy solution to
the field equation of
the O(3) model.  The function $\omega$ may in fact be meromorphic, but usually
the following
entire function is used:
\beq{240}
\omega(z) = \prod_n\left(\frac{z-z_{0n}}{\rho_n}\right)^{q_n}
\eeq
where the $\rho$'s are real constants, the $z_0$'s are complex constants, and
the $q$'s are
positive integers.  Each of the factors in the product represents a `soliton'
centred on the
point $(x_{1n},x_{2n}) = (\Re z_{0n},\Im z_{0n})$ and having topological charge
$q_n$.  The
quantity $\rho_n$ is a scale factor which determines the size of the soliton.
Replacing $z$
with $z^*$ in \eqnum{240} gives the corresponding `antisoliton' solution
$\omega(z^*)$.  The
total topological charge $Q$ is independent of the $z_0$'s and $\rho$'s:
\beq{250}
Q = \mp\sum_n q_n.
\eeq
The solution which represents a single soliton of charge $q$, or a single
antisoliton of
charge $-q$, may always be brought to the form
\beq{260}
W = \left\{\begin{array}{ll} z^{*q} & \mbox{(antisoliton)} \\ z^q &
\mbox{(soliton)}
\end{array}\right.
\eeq
by rescaling the Cartesian coordinates $x_1$ and $x_2$.

A number of other solutions to the field equation \eqnum{90} have come to light
since Belavin
and Polyakov's work.  There is the so-called `meron'
solution\cite{gross78,dealfaroetal78}
\beq{270}
W = \frac{\omega}{|\omega|}
\eeq
which is analogous to the meron solution in four-dimensional Yang-Mills theory
\cite{callanetal77}.  The function $\omega$ in \eqnum{270} is still the
analytic BP solution
but the total energy of a meron is infinite, which is expected in the light of
the
uniqueness proof of Garber {\it et al.}\cite{garberetal79}  Then a solution was
found which
contains both
\eqnum{230} and \eqnum{270} as special cases, and which in fact interpolates
between them
\cite{takhom81,tseitlin83}.  This solution can be written as
\beq{280}
W = R\E^{\I\Phi}
\eeq
where
\bsub{290}
\beq{290a}
R^2 = \frac{1 + D\,\sn(\eta|D)}{1 - D\,\sn(\eta|D)}
\eeq
and
\beq{290b}
\eta \equiv \ln|\omega| \comma \Phi =
-\I\ln\left(\frac{\omega}{|\omega|}\right).
\eeq
\esub
In these expressions, $\omega$ is the BP solution, $D$ is a real parameter
lying the range
$0 \leq D \leq 1$, and $\sn(\eta|D)$ is a Jacobi elliptic
function\cite{abrasteg}
of the quantity $\eta$ with parameter $D$.  When $D=1$ the solution \eqnum{290}
reduces to
\eqnum{230}, and when $D=0$ it reduces to \eqnum{270}.  If the function
$\omega$ is analytic
in $z$ then \eqnum{290} interpolates between solitons and merons, whereas if
$\omega$ is
analytic in $z^*$ then the interpolation is between antisolitons and
antimerons.  Some
special cases of \eqnum{290} where the function $\omega$ took certain forms,
such as
\eqnum{240}, have been examined\cite{khare80,ghikavis82,abbott82}.

A yet more general solution to \eqnum{90} was discovered by Purkait and
Ray\cite{purkaitray86}.  They took
the modulus $R=R(U)$ and argument $\Phi=\Phi(U,V)$ of $W$ to be functions of
the real and
imaginary parts $U(x_1,x_2)$ and $V(x_1,x_2)$ of an analytic function $T=U+\I
V$.  By direct
substitution into the field equation, they found that $\Phi$ is given by
\bsub{300}\beq{300a}
\Phi(U,V) = c_2 V + c_3\int\frac{(R^2 + 1)^2}{R^2}\,\dee U
\eeq
and that $R(U)$ is given by inverting the elliptic integral
\beq{300b}
U(R) = \half\int\,[Q(x)]^{-\half}\,\dee x
\eeq
where $x \equiv R^2$ and
\beq{300c}
Q(x) \equiv -c_3^2 x^4 + (c_1 - 2c_3^2)x^3 + (2c_1 + c_2^2 - 2c_3^2)x^2 + (c_1
- 2c_3^2)x -
            c_3^2
\eeq\esub
with $c_1$, $c_2$, $c_3$ being arbitrary real constants.  There are two special
cases of this
solution which will concern us here.  One is the interpolating solution
\eqnum{290}
encountered above: the Purkait-Ray solution \eqnum{300} reduces to \eqnum{290}
if
\beq{330}
c_1 \neq 0 \comma c_2 \neq 0 \comma c_3 = 0 \comma D^2 \equiv 4c_1c_2^{-2} + 1
> 0
\eeq
and
\beq{340}
T \equiv \recip{c_2}\,\ln\omega \comma \mbox{\ie\hsp{2}} \eta \equiv c_2 U.
\eeq
The second case of interest is that where $c_1 \neq 0$, $c_2 \neq 0$, $c_3 \neq
0$.  In this
case we find\cite{ody93}
\bsub{350}
\beq{350a}
R^2 = \frac{1 + \beta\,\sn(\alpha\eta|\tw{D})}{1 -
\beta\,\sn(\alpha\eta|\tw{D})}
\eeq
\beq{350b}
\Phi = c_2 V + \frac{4c_3}{c_2}\int\frac{\dee\eta}{1 -
\beta^2\,\sn^2(\alpha\eta|\tw{D})}
\eeq
\esub
where $\eta \equiv c_2 U$ and
\beq{355}
\tw{D} \equiv \frac{\beta}{\alpha}
\eeq
\beq{360}
\alpha^2 \equiv \frac{c_1 - 6c_3^2 + \delta}{c_1 + 2c_3^2 + \delta} \comma
\beta^2 \equiv \frac{c_1 - 6c_3^2 - \delta}{c_1 + 2c_3^2 - \delta}
\eeq
\beq{370}
\delta^2 \equiv (c_1 + 2c_3^2)^2 + 4c_2^2 c_3^2
\eeq
where the constants $c_1$, $c_2$, $c_3$ are such that $\alpha^2 > 0$ and
$\beta^2 > 0$.  This
second case of interest will be referred to as the `helicoidal solution', for
reasons to be
seen below.  Note that this solution is separate from the interpolating
solution \eqnum{290},
\ie \eqnum{350} does not reduce to \eqnum{290} if $c_3$ is set to zero.  This
is because if
$c_3=0$ then $\delta=\pm c_1$, and therefore either $\alpha$ or $\beta$ is
indeterminate.

\newpage
\section{Surfaces of constant mean curvature}
\label{geom}

In this section we summarise some features of the classical differential
geometry of surfaces
and, in particular, of surfaces of constant mean curvature.  For further
details, see refs.\
\citen{eisenhart} or \citen{struik}, for example.

A surface $S$ in $\R^3$ is uniquely specified, up to its overall position in
space, by giving
its first and second fundamental forms $I$ and $\II$:
\bsub{400}\bea
I &=& E(\dee u^2 + \dee v^2) \label{400a} \\
\II &=& L\,\dee u^2 + 2M\,\dee u\,\dee v + N\,\dee v^2 \label{400b}
\eea\esub
where $(u,v)$ are isothermal surface parameters and $E$, $L$, $M$, $N$ are
functions of $u$
and $v$, subject to certain constraints described below.  The form $I$ is just
the metric on
the surface $S$.  Conversely, if the position vector $\V{X}(u,v)$ of $S$ is
known then we can
find the four functions from
\bsub{410}\beq{410a}
E = |\V{X}_u|^2 = |\V{X}_v|^2
\eeq
\beq{410b}
L = -\V{X}_u\cdot\V{n}_{,u} \comma
M = -\V{X}_u\cdot\V{n}_{,v} \comma
N = -\V{X}_v\cdot\V{n}_{,v}
\eeq\esub
where $\V{X}_u \equiv \V{X}_{,u} \equiv \p_u\V{X}$ and $\V{X}_v \equiv
\V{X}_{,v} \equiv
\p_v\V{X}$ are the two tangent vectors to the surface and
\beq{420}
\V{n} \equiv E^{-1}(\V{X}_u \times \V{X}_v)
\eeq
is the unit normal vector.  We now introduce the mean curvature $H(u,v)$ and
Hopf function
\cite{hopf83} $\Psi(\zeta,\zeta^*)$ of the surface:
\beq{430}
H = {\ts\half}\E^{-2\Sigma}(L+N)
\eeq
\beq{440}
\Psi = -2\,\pdiff{\V{X}}{\zeta}\cdot\pdiff{\V{n}}{\zeta} = {\ts\half}(L-N) - \I
M,
\eeq
where
\beq{450}
E \equiv \E^{2\Sigma}
\eeq
and
\beq{460}
\zeta \equiv u + \I v.
\eeq
These functions $H$ and $\Psi$ encode many properties of the surface $S$.  In
terms of these
quantities, the forms $I$ and $\II$ may be written
\bsub{470}\bea
I &=& \E^{2\Sigma}|\dee\zeta|^2 \label{470a} \\
\II &=& {\ts\half}(\Psi\,\dee\zeta^2 + \Psi^*\dee\zeta^{*2}) +
H\E^{2\Sigma}|\dee\zeta|^2
        \label{470b}
\eea\esub
\ie
\bsub{480}\bea
L &=& HE + \Re\Psi \label{480a} \\
M &=& -\Im\Psi \label{480b} \\
N &=& HE - \Re\Psi. \label{480c}
\eea\esub
If $E$, $L$, $M$, $N$ are {\it given} functions then the position vector
$\V{X}(u,v)$ may in
principle be found by solving the Gauss-Weingarten equations.  In isothermal
coordinates
these equations take the form
\bsub{490}\bea
\V{X}_{u,u} &=& \Sigma_{,u}\V{X}_u - \Sigma_{,v}\V{X}_v + L\V{n} \label{490a}
\\
\V{X}_{u,v} &=& \Sigma_{,v}\V{X}_u + \Sigma_{,u}\V{X}_v + M\V{n} \label{490b}\\
\V{X}_{v,v} &=& -\Sigma_{,u}\V{X}_u + \Sigma_{,v}\V{X}_v + N\V{n} \label{490c}
\eea\esub\vsp{-14}\bsub{500}\bea
\msp{2.4}\V{n}_{,u} &=& -\E^{-2\Sigma}(L\V{X}_u + M\V{X}_v) \label{500a} \\
\V{n}_{,v} &=& -\E^{-2\Sigma}(M\V{X}_u + N\V{X}_v) \label{500b}
\eea\esub
These over-determined linear equations are subject to certain non-linear
compatibility
conditions which follow from the requirement that mixed second-order
derivatives of
$\V{X}_u$, $\V{X}_v$ and $\V{n}$ commute.  These are the Gauss and the
Codazzi-Mainardi
conditions, and in isothermal coordinates they take the respective forms
\beq{510}
LN - M^2 = -\E^{2\Sigma}\del^2\Sigma
\eeq
\beq{520}
\pdiff{\Psi}{\zeta^*} - \E^{2\Sigma}\pdiff{H}{\zeta} = 0.
\eeq
Another important quantity is the Gaussian curvature $K(u,v)$:
\beq{530}
K = \frac{LN - M^2}{E^2} = -\E^{-2\Sigma}\del^2\Sigma.
\eeq
If $K$ is integrated over the whole surface area $A$ of $S$ then the result is
the total
curvature, denoted $K_{\rm tot}$:
\beq{540}
K_{\rm tot} \equiv \int_S K\,\dee A = \int_S KE\,\dee u\wedge\dee v
		 = -\int_S\del^2\Sigma\;\dee u\wedge\dee v,
\eeq
and if $S$ is compact then it follows from the Gauss-Bonnet theorem that
\beq{550}
K_{\rm tot} = 2\pi\chi
\eeq
where $\chi$ is the Euler characteristic of $S$.  These relations will be
needed in the
remainder of this paper.

The Gauss map of a surface $S$ is constructed by parallel-transporting the
normal vector
$\V{n}$ at each point of $S$ to some origin $O$ in $\R^3$.  Since $\V{n}$ is a
unit vector,
the locus of the tips of these transported vectors will be some region $\Omega$
of the unit
sphere $\ess^2$.  The region $\Omega$ is called the spherical image of $S$.
The mapping
induced by $\V{n}$ from the surface $S$ to its spherical image is the Gauss
map:
\beq{560}
\V{n} : S \arrow \ess^2.
\eeq
The normal vector itself is sometimes referred to as `the Gauss map'.  It is
straightforward
to find the degree $\delta$ of the mapping \eqnum{560}: if $\V{n}^*\beta$ is
the pullback to
$S$ of the normalised element of area $\beta$ on $\ess^2$ then
\bea
\delta &=& \int_S \V{n}^*\beta \nonumber \\[2mm]
       &=& \recip{4\pi}\int_S \V{n}\cdot\V{n}_{,u}\times\V{n}_{,v}\,\dee
u\wedge\dee v
           \nonumber \\[2mm]
       &=& \recip{4\pi}\int_S KE\,\dee u\wedge\dee v \nonumber \\[2mm]
       &=& \frac{K_{\rm tot}}{4\pi} \label{570} \\[2mm]
       &=& \frac{\chi}{2} \label{580}
\eea
where \eqnum{580} holds if the surface is compact.  Two kinds of surface which
are important
to this work are spheres and minimal surfaces.  These surfaces have the
property that they are conformal to their spherical images\cite{eisenhart}.

We shall now describe these, and other, surfaces of constant mean curvature in
more detail.
In \Sec{main} it will be seen how they correspond to the various
time-independent solutions
to the O(3) sigma model described in \Sec{model}.  Note first that if the mean
curvature $H$
of a surface $S$ is a constant then the Codazzi-Mainardi condition \eqnum{520}
implies that
the Hopf function $\Psi$ is a function of $\zeta$ only:
\beq{600}
\Psi = \Psi(\zeta) \msp{8}\Leftrightarrow\msp{8} \mbox{$S$ is a surface of
constant mean
curvature (CMC).}
\eeq

\subsection*{The sphere}

Consider first a CMC surface having $\Psi = 0$, $H \neq 0$.  In this case, the
first and
second fundamental forms \eqnum{470} are related by
\beq{610}
\II = HI,
\eeq
from which it follows \cite{struik} that $S$ is a sphere of radius $H^{-1}$.
Since $H$ is
therefore simply an overall scale factor for the surface, we may set $H = 1$
without loss of
generality.
\beq{620}
\Psi = 0, \msp{2} H = 1 \msp{8}\Leftrightarrow\msp{8} \mbox{$S$ is a sphere of
unit radius.}
\eeq

\subsection*{Minimal surfaces}

Now let the mean curvature $H$ be zero at all points $(u,v)$ of the surface
$S$.  This is the
defining property of a minimal surface \cite{nitsche}.  The Hopf function
$\Psi(\zeta)$
determines a minimal surface via the Weierstrass-Enneper representation formula
for the
position vector $\V{X}(u,v)$:
\bsub{630}
\beq{630a}
\V{X}(u,v) = \half\,\Re\!\int\V{V}(\zeta)\Psi(\zeta)\,\dee\zeta
\eeq
where the vector $\V{V}(\zeta)$ is
\beq{630b}
\V{V}(\zeta) \equiv \bve \zeta^2 - 1 \\ \I(\zeta^2 + 1) \\ 2\zeta \eve.
\eeq\esub
To each analytic function $\Psi(\zeta)$ there corresponds a minimal surface $S$
given by
\eqnum{630}.  From \eqnum{630} it is straightforward to derive the forms $I$
and $\II$ for a
minimal surface:
\bsub{640}\bea
  I &=& {\ts\quar}|\Psi|^2(1+|\zeta|^2)^2|\dee\zeta|^2 \label{640a} \\
\II &=& {\ts\half}(\Psi\,\dee\zeta^2 + \Psi^*\dee\zeta^{*2}) \label{640b}
\eea\esub
\beq{650}
\mbox{\hspace{-\mathindent}\ie} \msp{5} L = -N = \Re\Psi \comma M = -\Im\Psi.
\eeq
The normal vector $\V{n}(u,v)$ is found to be
\beq{660}
\V{n} = \recip{1+|\zeta|^2}\bve 2\,\Re\zeta \\ -2\,\Im\zeta \\ 1-|\zeta|^2 \eve
\eeq
which is notable because it does not depend on the Hopf function $\Psi(\zeta)$.
 Therefore,
locally, any two minimal surfaces have the same Gauss map.  In this sense
minimal surfaces
are exceptional because in general a surface is essentially determined by its
Gauss map
\cite{hoffoss85}.  The form \eqnum{660} of $\V{n}$ indicates that $\zeta$ can
be regarded as
a stereographic projection of the Gauss map.  The complex variable $\zeta = u +
\I v$ itself
will therefore sometimes be referred to as the Gauss map of the surface.

The metric \eqnum{640a} on a minimal surface depends on the {\it modulus} of
the Hopf
function $\Psi(\zeta)$.  Therefore a one-parameter family of isometric minimal
surfaces may
be obtained by subjecting $\Psi$ to the Bonnet transformation
\beq{670}
\Psi \,\arrow \E^{\I\alpha}\Psi \comma 0 \leq \alpha < 2\pi
\eeq
where $\alpha$ is the `Bonnet angle'.  The surfaces obtained in this way from a
given minimal
surface $S$ are said to be associate to $S$.  On the other hand, if $\Psi$ is
multiplied by a
{\it real} constant $c$ then the surface $S$ is simply scaled in size by a
factor $c$.  Such
a transformation does not give rise to an essentially distinct surface and is
therefore
trivial.

The simplest choice for $\Psi$, namely
\beq{680}
\Psi = 1 \comma
\eeq
gives rise to Enneper's minimal surface \cite{nitsche}.  This surface is an
example of a Bour
surface: a minimal surface which is applicable (\ie isometric via a
one-parameter
transformation) to a surface of revolution.  The Hopf function $\Psi(\zeta)$ of
an arbitrary
Bour surface is given by \cite{eisenhart}
\beq{690}
\Psi = c\msp{0.2}\E^{\I\alpha}\zeta^{b-2}
\eeq
where $c\msp{0.2}\E^{\I\alpha}$ is an arbitrary complex number incorporating
the scale
factor $c$ and Bonnet angle $\alpha$, and $b$ is an arbitrary real constant
which determines
the Bour surface.  Specifying $b$ fixes the `Bonnet class' of the surface, \ie
the family of
associate minimal surfaces to which it belongs, and the angle $\alpha$
determines a unique
surface in that class.

If a coordinate transformation
\beq{700}
u = u(x_1,x_2) \comma v = v(x_1,x_2)
\eeq
to new independent parameters $(x_1,x_2)$ is made then the Gauss map $\zeta$
will be regarded
as a complex function of the variables $z \equiv x_1 + \I x_2$ and $z^*$:
\beq{710}
\zeta = \zeta(z,z^*).
\eeq
If $\zeta$ is the Gauss map of a minimal surface or a sphere then it is an {\it
analytic}
function of $z$ or $z^*$ respectively.  Similarly, if $\zeta(z,z^*)$ is the
Gauss map of a
CMC surface (with $H \neq 0$) then it is {\it harmonic} \cite{ruhvilms70}.  If,
in addition,
both coordinate systems $(u,v)$ and $(x_1,x_2)$ are isothermal then the Hopf
function
$\psi$ in the new parametrisation is related to $\Psi$ by \cite{hopf83}
\beq{720}
\Psi\,\dee\zeta^2 = \psi\,\dee z^2.
\eeq
Our use of lower case to denote the Hopf function in the
$(x_1,x_2)$-parametrisation will be
applied to the other quantities of surface theory.  When referred to the
parameters $u$ and
$v$ we will write, for example, $E$, $L$, $M$, $N$, $\Sigma$, $\Psi$, {\it
etc}.  The
corresponding quantities in the $(x_1,x_2)$-parametrisation will be written
$e$, $l$, $m$,
$n$, $\sigma$, $\psi$, {\it etc}.  The notation of quantities which are
invariant under a
transformation of the form \eqnum{700}, such as $K$, $H$, $\V{n}$, {\it etc.},
will not be
changed\fn{$H$ and $\V{n}$, for example, are invariant up to a sign.  If the
parameter
transformation preserves the orientation of the surface then the sign of these
quantities
does not change.}\hsp{2.5}.

Many general results pertaining to arbitrary CMC surfaces, and, indeed, to
surfaces in
general may be obtained from the Kenmotsu representation
formula\cite{kenmotsu79} which gives
the position vector $\V{X}(x_1,x_2)$ of a surface in terms of its mean
curvature $H(x_1,x_2)$
and Gauss map $\zeta(z,z^*)$.  For example, we find for a CMC surface the
following
relations:
\beq{730}
I = \E^{2\sigma}|\dee z|^2 \mbox{\hsp{5} where \hsp{2}}
\E^{2\sigma} = \frac{4|\p^*\zeta|^2}{H^2(1 + |\zeta|^2)^2}
\eeq
\beq{740}
\psi(z) = \E^{2\sigma}H\frac{\p\zeta}{\p^*\zeta}
\eeq
\beq{750}
K_{\rm tot} = 4\int_S\frac{|\p^*\zeta|^2 - |\p\zeta|^2}{(1 +
|\zeta|^2)^2}\,\dee x_1 \wedge
              \dee x_2.
\eeq
The Gauss map $\zeta(z,z^*)$ is harmonic, \ie it satisfies
\beq{760}
\p\p^*\zeta - \frac{2\zeta^*\,\p\zeta\,\p^*\zeta}{1 + |\zeta|^2} = 0,
\eeq
and the normal vector $\V{n}(x_1,x_2)$ is related to $\zeta$ by
\beq{765}
\V{n} = \recip{1+|\zeta|^2}\bve 2\,\Re\zeta \\ -2\,\Im\zeta \\ 1-|\zeta|^2
\eve.
\eeq

An important result in the theory of CMC surfaces is the following: the
directions of
principle curvature on a CMC surface form an isothermal coordinate system
\cite{eisenhart}.
An immediate consequence is that, in a parametrisation by these lines of
curvature, the
second fundamental form $\II$ is diagonal and therefore that $\psi(z)$ is
purely real:
\beq{770}
m = -\Im\psi = 0.
\eeq
Moreover, it follows from the analyticity of $\psi$\,---\,see
\eqnum{600}\,---\,that $\psi$
is a real constant.

There are two further kinds of CMC surfaces which are relevant to the O(3)
sigma model and
these are described next.

\subsection*{Delaunay surfaces}

The sphere, for example, is a surface of revolution of constant mean curvature,
or
Delaunay surface\cite{eells87}.  Delaunay surfaces form a one-parameter family
for which a
general expression for the position vector $\V{X}$ exists \cite{kenmotsu80}.
Let $C =
\{(X(s),Z(s))\}$ be a curve in the $(xz)$-plane of $\R^3$, parametrised by
arc-length $s$.
Revolving the curve $C$ around the $z$-axis leads to a surface of revolution
$S$ with
position vector
\beq{780}
\V{X}(s,\theta) = \bve X(s)\cos\theta \\ X(s)\sin\theta \\ Z(s) \eve
\eeq
where $0 \leq \theta < 2\pi$.  From \eqnum{780} we find
\bsub{790}\bea
I &=& \dee s^2 + X^2\,\dee\theta^2 \label{790a} \\
\II &=& (X'Z'' - X''Z')\dee s^2 + XZ'\,\dee\theta^2 \label{790b}
\eea\esub
which shows that the parametric lines of $S$, \ie its meridians and parallels,
are also its
lines of principal curvature (because $\II$ is diagonal).  However, the
parameters $(s,\theta)$ are not
isothermal; an isothermal coordinate system will be introduced below.  The
functions $X(s)$
and $Z(s)$ can be found explicitly once the mean curvature $H(s,\theta)$ is
given
\cite{kenmotsu80}.  In the present case we require $H =$ constant, and it
proves convenient
to choose $H = \half$.  Then it can be shown that
\beq{800}
X = (1 + D^2 + 2D\sin s)^\half
\eeq
\beq{810}
Z = \int\frac{1 + D\sin s}{X}\,\dee s
\eeq
where $D \geq 0$ is a real parameter.  The surface corresponding to $D=0$ is
the right-%
circular cylinder; when $0<D<1$ the Delaunay surface is an unduloid; when $D=1$
it is a
sphere; and for $D>1$ the surface is called a nodoid.

To make use of the general results for CMC surfaces described above an
isothermal
parametrisation of the Delaunay surfaces must be introduced.  Defining the new
variable
$\eta(s)$ by
\beq{820}
\eta \equiv \int\frac{\dee s}{X}
\eeq
causes the first fundamental form \eqnum{790a} to become
\beq{830}
I = X^2(\dee \eta^2 + \dee\theta^2).
\eeq
Therefore the parameters $(\eta,\theta)$ are isothermal.  The problem now is to
find the
Gauss map $\zeta$ of a Delaunay surface, and it is shown in the Appendix that
\beq{840}
\zeta = R\E^{-\I\Phi}
\eeq
where
\bsub{850}
\beq{850a}
R^2 = \frac{1 + D\,\sn(\eta|D)}{1 - D\,\sn(\eta|D)}
\eeq
\beq{850b}
\Phi = \theta + \pi.
\eeq\esub
Now, it is a fact \cite{struik} that two isothermal parametrisations
$(\eta,\theta)$ and
$(x_1,x_2)$, say, of a surface $S$ are related by
\beq{860}
\eta + \I\theta = \tau(x_1 \pm \I x_2)
\eeq
where $\tau$ is a complex analytic function of either $z \equiv x_1 + \I x_2$
or $z^*$.  In
the present case, where $S$ is a Delaunay surface, the function $\tau$ must be
analytic in
$z$.  To see why this is so, consider the special case $D = 1$.  The surface is
then a
sphere.  When $D = 1$ the elliptic function $\sn(\eta|D)$ becomes
\beq{870}
\sn(\eta|1) = \tanh\eta
\eeq
and therefore the Gauss map \eqnum{850} reduces to
\beq{880}
\zeta = \E^{\eta - \I\Phi} = -\E^{\tau^*}.
\eeq
But the Gauss map of a sphere is analytic in $z^*$.  Consequently, \eqnum{880}
implies that
\beq{890}
\tau = \tau(z).
\eeq
It follows from Kenmotsu's representation theorem for CMC surfaces that if
$\eta(x_1,x_2)$
and $\theta(x_1,x_2)$ are such that \eqnum{890} holds then the function
$\zeta(z,z^*)$ given
by \eqnum{840} and \eqnum{850} is harmonic.  It will be seen presently that
this is indeed
the case.

\subsection*{Helicoids of constant mean curvature}

A natural generalisation of a surface of revolution is the surface formed when
the generating
curve $C$ is simultaneously revolved around the $z$-axis and translated, at
constant speed,
in a direction parallel to the $z$-axis.  Such a surface is called a
generalised helicoid.
Let $2\pi h$, where $h$ is a constant, be the distance in the $z$-direction by
which a given
point on $C$ moves after one revolution of $C$ about the $z$-axis.  Then the
position vector
of a generalised helicoid is given by
\beq{900}
\V{X}(s,\theta) = \bve X(s)\cos\Delta(s,\theta) \\
		       X(s)\sin\Delta(s,\theta) \\
		       Z(s) + h\Delta(s,\theta) \eve
\eeq
where $X(s)$ and $Z(s)$ define the curve $C$ and $\Delta(s,\theta)$ is a
function to be
found.	Generalised helicoids of constant mean curvature were first studied by
do Carmo and
Dajczer\cite{docdaj82}.  If the mean curvature is chosen to be $H = \half$ then
their
expressions for the functions $X$, $Z$ and $\Delta$ take the forms
\beq{910}
X = (1 + D^2 + 2D\sin s)^\half
\eeq
\beq{920}
Z = \int\frac{Y(1 + D\sin s)}{X^2}\,\dee s
\eeq
\beq{930}
\Delta = \theta - h\int\frac{1 + D\sin s}{X^2 Y}\,\dee s
\eeq
where
\beq{940}
Y^2 \equiv X^2 + h^2.
\eeq
Clearly, the Delaunay surfaces are obtained if $h$ is set to zero.

The first fundamental form of a CMC helicoid is
\beq{950}
I = \dee s^2 + Y^2\,\dee\theta^2 \comma
\eeq
which may be brought to the isothermal form
\beq{960}
I = Y^2(\dee\eta^2 + \dee\theta^2)
\eeq
by the introduction of the new variable
\beq{970}
\eta \equiv \int\frac{\dee s}{Y}.
\eeq
Again the quantity of interest is the Gauss map $\zeta$.  It is shown in the
Appendix that
$\zeta$ is given by \eqnum{840} where now
\bsub{980}
\beq{980a}
R^2 = \frac{1 + \beta\,\sn(\alpha\eta|\tw{D})}{1 -
\beta\,\sn(\alpha\eta|\tw{D})}
\eeq
\beq{980b}
\Phi = \theta + \pi - h\int\frac{\dee\eta}{1 -
\beta^2\,\sn^2(\alpha\eta|\tw{D})}
\eeq\esub
with
\bsub{990}
\beq{990a}
\tw{D} = \frac{\beta}{\alpha}
\eeq
\beq{990b}
\alpha \equiv {\ts\half}(A + B) \comma \beta \equiv {\ts\half}(A - B)
\eeq
\beq{990c}
A^2 \equiv (1 + D)^2 + h^2 \comma B^2 \equiv (1 - D)^2 + h^2.
\eeq\esub
Just as for Delaunay surfaces, if the isothermal parameters $(\eta,\theta)$ are
related to
other isothermal coordinates $(x_1,x_2)$ by \eqnum{890} then the Gauss map
$\zeta(z,z^*)$
will be harmonic.

\newpage
\section{CMC surfaces and the O(3) model}
\label{main}

Now we combine the material of the previous two sections to give an explicit
correspondence
between static solutions to the non-linear O(3) sigma model and surfaces of
constant mean
curvature in $\R^3$.

The correspondence is founded upon the identification of the field vector
$\V{n}$ of the O(3)
model with the normal vector, or, more correctly, the Gauss map, of a surface.
Both objects
are three-component two-parameter unit vectors taking values in the subset
$\ess^2$ of
$\R^3$.	 The explicit functional form of the relation \eqnum{700} between the
spatial
coordinates $(x_1,x_2)$ and the general surface parameters $(u,v)$ is to be
found.

The starting point is to combine the duality or field equation of the O(3)
model with the
differential equations of surface theory.  If the field vector and normal
vector are to be
identified then both sets of equations must be satisfied simultaneously.  By
adhering to the
notational convention described in \Sec{geom} the fundamental forms $I$ and
$\II$ of the
surfaces to be found will be denoted
\bsub{1400}\bea
I &=& E(\dee x_1^2 + \dee x_2^2) = \E^{2\sigma}|\dee z|^2 \label{1400a} \\
\II &=& l\,\dee x_1^2 + 2m\,\dee x_1\,\dee x_2 + n\,\dee x_2^2
      = {\ts\half}(\psi\,\dee z^2 + \psi^*\dee z^{*2}) + H\E^{2\sigma}|\dee
z|^2
        \label{1400b}
\eea\esub
where $H(x_1,x_2)$ and $\psi(z,z^*)$ are the mean curvature and Hopf function,
respectively,
in the $(x_1,x_2)$-parametrisation:
\beq{1410}
H = {\ts\half}\E^{-2\sigma}(l+n)
\eeq
\beq{1420}
\psi = {\ts\half}(l-n) - \I m.
\eeq
In terms of $x_1$ and $x_2$ the Weingarten equations \eqnum{500} are
\bsub{1430}\bea
\V{n}_{,1} &=& -\E^{-2\sigma}(l\V{X}_1 + m\V{X}_2) \label{1430a} \\
\V{n}_{,2} &=& -\E^{-2\sigma}(m\V{X}_1 + n\V{X}_2) \label{1430b}
\eea\esub
where $\V{X}_i \equiv \p_i\V{X}$.  By differentiating equations \eqnum{1430}
again it is
found that
\bea
\del^2\V{n} &=& -\E^{-2\sigma}\{[l_{,1} + m_{,2} - (l+n)\sigma_{,1}]\V{X}_1 +
\nonumber \\
& & \msp{12.5}+ [m_{,1} + n_{,2} - (l+n)\sigma_{,2}]\V{X}_2 + (l^2 + 2m^2 +
n^2)\V{n}\}.
\label{1440}
\eea
We can now relate some surface-geometric quantities to the energy density
$\ED(x_1,x_2)$ of
the O(3) model.  If either \eqnum{1430} or \eqnum{1440} is used in the
expression \eqnum{110}
for $\ED$ then we find
\bea
\ED &=& -\E^{-2\sigma}(l^2 + 2m^2 + n^2) \nonumber \\
    &=& \E^{2\sigma}H^2 + \E^{-2\sigma}|\psi|^2. \label{1460}
\eea
The Gaussian curvature $K(x_1,x_2)$ may be expressed in terms of $\ED$.  We
find from
\eqnum{1400} and \eqnum{1460} that
\bea
K &=& \E^{-4\sigma}(ln - m^2) \nonumber \\
  &=& \E^{-2\sigma}\ED - 2\E^{-4\sigma}|\psi|^2 \label{1470} \\
  &=& 2H^2 - \E^{-2\sigma}\ED. \label{1480}
\eea
These general relations will be applied to the special case of CMC surfaces
below.

We consider the two classes of solution to the O(3) model separately: those
which satisfy the
duality equation \eqnum{200}, and those which do not.

\subsection*{Surfaces arising from the duality equation}

In the duality equation \eqnum{200} we substitute for $\V{n}_{,1}$ and
$\V{n}_{,2}$ from
the Weingarten equations \eqnum{1430}, and for $\V{n}$ from
\beq{1490}
\V{n} = \E^{-2\sigma}(\V{X}_1 \times \V{X}_2).
\eeq
Then, on equating coefficients of $\V{X}_1$ and $\V{X}_2$, we find that the
duality equation
implies the following constraints on the second fundamental form:
\beq{1500}
n = \mp l \comma m = \pm m.
\eeq
We consider the two choices of sign separately.

The solutions to the O(3) model which follow from choosing the lower sign in
the duality
equation, and hence in \eqnum{1500}, are the Belavin-Polyakov solitons given by
\eqnum{230}
and \eqnum{240}.  The surfaces corresponding to these solutions are to be found
and will be
denoted $S_+$.  From \eqnum{1500} we find that the Hopf function \eqnum{1420}
of $S_+$ is
identically zero,
\beq{1510}
\psi = 0,
\eeq
and therefore from \eqnum{1400} that
\beq{1520}
\II = H\E^{2\sigma}|\dee z|^2 = HI
\eeq
where
\beq{1530}
H = \E^{-2\sigma}l.
\eeq
The Codazzi-Mainardi condition \eqnum{520}, which in the present notation reads
\beq{1540}
\p^*\psi - \E^{2\sigma}\p H = 0,
\eeq
therefore implies that $H$ is a constant, as a result of \eqnum{1510}.  Note
that $H$ cannot
be identically zero, for it would follow from \eqnum{1520} that $\II = 0$ and
therefore that
the surface $S_+$ is a plane.  A plane has a trivial Gauss map\,---\,a single
point on
$\ess^2$\,---\,which corresponds to the ground state $\V{n} = \V{n}_0$ of the
O(3) model.
Consequently, we shall assume here that $H =$ constant $\neq 0$.  We can now
use results from
the theory of CMC surfaces.  The mean curvature $H$ is just a scale factor for
such a surface
and may be set to unity, for convenience.  Then we see from \eqnum{1520} that
$I = \II$.
Thus, from \eqnum{620}, the surface $S_+$ corresponding to soliton solutions of
the O(3)
model is the unit sphere $\ess^2$.  This result is somewhat surprising when one
considers the
multitude of possible functional forms of the field vector $\V{n}(x_1,x_2)$
which can follow
from the general soliton configuration \eqnum{240}.  Yet the surface to which
$\V{n}$ is
normal is always the unit sphere.  But, clearly, different functions
$\omega(z)$ will lead to
different parametrisations of the sphere $S_+$.  This can be made explicit by
showing that
$I$ can always be brought to the standard form of the metric on a sphere.  To
see this,
consider the expression \eqnum{1460} for the energy density $\ED$.  We find
immediately that
\beq{1550}
\E^{2\sigma} = \ED.
\eeq
But the energy density $\ED$ of a soliton configuration $W = \omega(z)$ is
found from
\eqnum{170} to be
\bea
\ED &=& \frac{4}{(1 + |\omega|^2)^2}\left|\diff{\omega}{z}\right|^2 \\[2mm]
    &=& \frac{\sin^2\!\Theta}{|\omega|^2}\,\frac{\dee |\omega|^2 +
|\omega|^2\dee\Phi^2}
	{|\dee z|^2} \label{1560}
\eea
where $\Theta$ and $\Phi$ are defined by \eqnum{80}.  Noting that
\beq{1570}
\diff{\Theta}{|\omega|} = \frac{\sin\Theta}{|\omega|}
\eeq
enables \eqnum{1550} and \eqnum{1560} to be combined to give
\beq{1580}
I = \ED|\dee z|^2 = \dee\Theta^2 + \sin^2\!\Theta\,\dee\Phi^2,
\eeq
as claimed.  The dependence of $\Theta$ and $\Phi$ on the variables $x_1$ and
$x_2$ will of
course depend on the functional form of $\omega(z)$.

We may find the explicit form of the relation \eqnum{700} between $(x_1,x_2)$
and $(u,v)$.
We equate expression \eqnum{765} for the normal vector of a CMC surface with
\eqnum{70} for
the field vector of the O(3) model.  This yields
\[ u(x_1,x_2) = \Re\,\omega(z) \comma v(x_1,x_2) = -\Im\,\omega(z) \comma \]
\ie the Gauss map $\zeta$ of $S_+$ is given by
\beq{1590}
\zeta(z^*) = \omega^*(z^*).
\eeq

Now consider the Gaussian curvature $K$ of the surface $S_+$.  We see
immediately from either
\eqnum{1470} or \eqnum{1480} that
\beq{1600}
K = 1,
\eeq
as expected for a unit sphere.	However, the {\it total} curvature $K_{\rm tot}$
is not
always $4\pi$.	It is in fact equal to the total energy $E_{\rm tot}$ of the
soliton
configuration $\omega(z)$:
\beq{1610}
K_{\rm tot} = \int K\E^{2\sigma}\,\dee x_1 \wedge \dee x_2
	    = \int\ED\,\dee x_1 \wedge \dee x_2 = E_{\rm tot} = 4\pi Q
\eeq
where $Q>0$.  Equation \eqnum{570} then confirms that $Q$ can be identified
with the degree
$\delta$ of the Gauss map of the surface.  A sphere is a compact surface and
therefore the
Euler characteristic $\chi$ is defined.  Invoking \eqnum{550} gives
\beq{1620}
\chi = 2Q.
\eeq
These facts show that when $Q>1$ the surface $S_+$ is actually a multiple
covering of the
unit sphere $\ess^2$.

Now consider the upper sign in the duality equation \eqnum{200}.  This choice
leads to the
Belavin-Polyakov antisolitons given by $W = \omega(z^*)$.  The surfaces to be
found in this
case will be denoted $S_-$.  Taking the upper sign in \eqnum{1500} gives
\beq{1625}
n = -l
\eeq
which shows straight away that the mean curvature \eqnum{1410} is identically
zero. Therefore
the surfaces $S_-$ corresponding to antisoliton solutions of the O(3) model are
minimal
surfaces.  The position vector of a minimal surface is given by the
Weierstrass-Enneper
representation formula \eqnum{630}.  The task is to find the functional form of
the Hopf
function $\psi$ which follows from the antisoliton configuration $\omega(z^*)$.
 Then the
minimal surfaces $S_-$ may be constructed explicitly.  To begin with, the
expressions for the
field vector and normal vector are equated to give
\beq{1630}
\zeta(z) = \omega^*(z).
\eeq
The metric factor $\E^{2\sigma}$ is found from \eqnum{1460} to be related to
the energy
density $\ED$ by
\beq{1640}
\E^{2\sigma} = |\psi|^2 \ED^{-1}.
\eeq
In order to make the results obtained in this section directly comparable with
those found in
the case of the lower sign, we shall take $x_1$ and $x_2$ to be parameters
along the lines of
curvature of $S_-$.  The Hopf function $\psi$ will then be a real constant
because $S_-$ is a
CMC surface.  So although taking the upper sign in \eqnum{1500} gives no
restriction on the
coefficient $m$, the parametrisation of $S_-$ may nevertheless be chosen to
make $m$ vanish.
For convenience we will set
\beq{1650}
\psi = 1.
\eeq
Then we find the fundamental forms \eqnum{1400} of $S_-$ to be
\bsub{1660}\bea
I &=& \ED^{-1}|\dee z|^2
    = {\ts\quar}(1 + |\zeta|^2)^2\left|\diff{z}{\zeta}\right|^2|\dee z|^2
\label{1660a}
      \\[2mm]
\II &=& {\ts\half}(\dee z^2 + \dee z^{*2}) \label{1660b}
\eea\esub
where \eqnum{1630} has been used in \eqnum{1660a}.  The function $z(\zeta)$ is
the inverse of
$\zeta(z)$.  Now, rearranging \eqnum{720} gives the form of the Hopf function
$\Psi(\zeta)$
in the $(u,v)$ parametrisation:
\beq{1670}
\Psi = \left(\diff{z}{\zeta}\right)^2.
\eeq
The Weierstrass-Enneper formula \eqnum{630} then gives the position vector
$\V{X}(u,v)$ of
the minimal surface $S_-$ corresponding to the analytic function $\Psi(\zeta)$.
 Let us
consider some examples.  A single antisoliton of unit topological charge is
represented by
the analytic function $\omega(z^*) = z^*$.  We find, then, that $\zeta =
\omega^* = z$ and
therefore from \eqnum{1670} that $\Psi = 1$.  The minimal surface defined by
this Hopf
function is Enneper's surface.  Next, consider a single antisoliton of
arbitrary charge $q$,
represented by the function \eqnum{260}:
\beq{1680}
\omega(z^*) = z^{*q}.
\eeq
The inverse function $z(\zeta)$ is clearly
\beq{1690}
z = \E^{2\pi\I s/q}\msp{0.2}\zeta^{1/q}
\eeq
where $s$ is any integer.  The Hopf function \eqnum{1670} is therefore
\beq{1700}
\Psi = \frac{1}{q^2}\,\E^{4\pi\I s/q}\msp{0.2}\zeta^{2/q - 2}.
\eeq
This is the Hopf function of a Bour surface: a comparison with \eqnum{690}
yields
\beq{1710}
b = \frac{2}{q} \comma \alpha = \frac{4\pi s}{q} \comma c = \recip{q^2}.
\eeq
The scale factor $c$ may be neglected; the Bour surface is determined
essentially by the
exponent $b$ and the Bonnet angle $\alpha$.  But note that fixing the value of
$q$ does not
specify $\alpha$ uniquely.  A moment's thought shows that, for a given value of
$q$, distinct
surfaces are generated if the integer $s$ takes the values
\beq{1720}
s = 1, 2,\,\ldots,\,d
\eeq
where the `multiplicity' $d$ is given by
\beq{1730}
d = \left\{\begin{array}{ll} q, & \mbox{if $q$ is odd,} \\ {\ds\frac{q}{2}}, &
\mbox{if $q$
is even.} \end{array}\right.
\eeq
Therefore there are $d$ Bour surfaces, associate to each other, corresponding
to an
antisoliton of charge $q$.

Calculations analogous to those in the case of the lower sign may be carried
out.  The
Gaussian curvature is found from \eqnum{1480} to be
\beq{1740}
K = -\E^{-4\sigma}
\eeq
and the total curvature $K_{\rm tot}$ to be
\beq{1750}
K_{\rm tot} = \int K\E^{2\sigma}\,\dee x_1 \wedge \dee x_2
	    = -\int\ED\,\dee x_1 \wedge \dee x_2 = -E_{\rm tot} = 4\pi Q
\eeq
where $Q<0$.  The integer $Q$ is again identified with the degree $\delta$ of
the Gauss map
but since there are no compact minimal surfaces the Euler characteristic $\chi$
is not
defined.

To summarise, finite-energy solutions to the two-dimensional O(3) sigma model
are induced by
the Gauss maps of surfaces in $\R^3$ which are conformal to their spherical
images.
Explicitly, the soliton solutions of the O(3) model are induced by the Gauss
maps
$\zeta(z^*)$ of multiple coverings of the unit sphere $\ess^2$, for which the
metric is given
by \eqnum{1580} and the mean curvature and Hopf function are
\beq{1760}
H = 1 \comma \psi = 0.
\eeq
Antisoliton solutions are induced by the Gauss maps $\zeta(z)$ of certain
minimal surfaces,
for which
\beq{1770}
H = 0 \comma \psi = 1.
\eeq
The minimal surfaces can be found explicitly by going to the
$(u,v)$-parametrisation, in
which the Hopf function $\Psi$ has the form \eqnum{1670}, and by using the
Weierstrass-%
Enneper representation formula \eqnum{630}.  In both cases, the functional form
of the Gauss
map $\zeta$ must be that of a polynomial in its argument.  Then a solution of
the form
\eqnum{240} will be induced.

\subsection*{Surfaces arising from the field equation}

We now turn to the general problem of determining which surfaces $S$ have
normal vectors
$\V{n}$ which obey the field equation, but not the duality equation, of the
O(3) sigma model.

The field equation \eqnum{40} implies simply that the coefficients of $\V{X}_1$
and $\V{X}_2$
in the expression \eqnum{1440} for $\del^2\V{n}$ must vanish.  By introducing
the mean
curvature $H$, given by \eqnum{1410}, these coefficients become
\bsub{1780}\bea
{\ts\half}(l-n)_{,1} + m_{,2} + \E^{2\sigma}H_{,1} &=& 0 \label{1780a} \\
{\ts\half}(l-n)_{,2} - m_{,1} - \E^{2\sigma}H_{,2} &=& 0. \label{1780b}
\eea\esub
Multiplying the second of these equations by i and adding it to the first gives
\beq{1790}
\p^*\psi + \E^{2\sigma}\p H = 0
\eeq
which may be combined with the Codazzi-Mainardi condition \eqnum{1540} to give
\beq{1800}
\p^*\psi = 0 \comma \p H = 0.
\eeq
These relations imply that the field equation of the O(3) model, when regarded
as a
constraint on the normal vector to a surface $S$, is the condition that $S$ be
of constant
mean curvature.  Now the two expressions \eqnum{70} and \eqnum{765} for the
vector $\V{n}$
may be equated.  The result is that the stereographic projection $\zeta(z,z^*)$
of the normal
vector can be identified in all cases with the complex conjugate of the
stereographic
projection $W(z,z^*)$ of the field vector:
\beq{1810}
\zeta(z,z^*) = W^*(z,z^*).
\eeq
This result encompasses \eqnum{1590} and \eqnum{1630}.  Also, the expressions
\eqnum{1610}
and \eqnum{1750} for the total curvature $K_{\rm tot}$ are special cases of the
following
result, obtained by comparing the integrand in \eqnum{750} with the charge
density
$J(x_1,x_2)$ given by \eqnum{180}:
\beq{1820}
K_{\rm tot} = 4\pi\int J\,\dee x_1 \wedge \dee x_2 = 4\pi Q.
\eeq
The relation $K_{\rm tot} = \mp E_{\rm tot}$ follows if the duality equation
holds, \ie in
the case of finite total energy.

Upon the identification of $\zeta$ with $W^*$, the Kenmotsu equation
\eqnum{760} becomes the
complex conjugate of the field equation \eqnum{90}.  Therefore any
time-independent solution
$W(z,z^*)$ of the O(3) model may be regarded as (the conjugate of) the Gauss
map of a CMC
surface $S$ in $\R^3$.  In principle, $S$ may be constructed explicitly by
inserting $\zeta =
W^*$ into the Kenmotsu representation formula \cite{kenmotsu79}.  Conversely,
every CMC
surface $S$ will induce via its Gauss map $\zeta$ a solution to the O(3) model.
 If $S$ is
minimal or a sphere then the proviso given above\,---\,that the Gauss map
$\zeta$ must be a
polynomial\,---\,must hold.  Examples of solutions induced by non-minimal,
non-spherical CMC
surfaces are given next.

\subsubsection*{The Delaunay solution}

The following result is an immediate consequence of identifying \eqnum{290} as
the Gauss map
\eqnum{850}: the Gauss map of a Delaunay surface induces the elliptic
interpolating solution
of the O(3) sigma model in the case where the function $\omega$ is analytic in
$z$.  The
constraint on $\omega$ is necessary for consistency with the special case $D =
1$, for in
that case the Delaunay surface is a sphere and it was found above that the
sphere $\ess^2$
corresponds to soliton solutions, for which $W = \omega(z)$.  Therefore, for
example, the
interpolation between a meron and a soliton is realised by the transformation
between a
cylinder ($D=0$) and a sphere ($D=1$).	 In this case, then, the interpolating
solution may
be dubbed the `Delaunay solution'.  On the other hand, if the function $\omega$
is analytic
in $z^*$ then when $D=1$ the interpolating solution \eqnum{290} reduces to the
antisoliton
configuration $W = \omega(z^*)$.  The surfaces which correspond to this
solution are the
minimal surfaces found above, none of which are contained in the one-parameter
family of
equations \eqnum{840} and \eqnum{850}.  So when $\omega = \omega(z^*)$ there is
an
obstruction to the interpretation of \eqnum{290} in terms of the Gauss map of
Delaunay
surfaces.  Instead, the interpolation between an antimeron and an antisoliton
corresponds to
a transformation between a cylinder and a minimal surface, but we know of no
family of CMC
surfaces which realises this.

If it is assumed that $\omega$ is analytic in $z$ then the following relation
between the
isothermal parameters $(\eta,\theta)$ on the Delaunay surface and the complex
function
$\omega(z)$ may be deduced from \eqnum{340}:
\beq{1830}
\omega = -\E^{\tau}
\eeq
where $\tau \equiv \eta + \I\theta$.  Equation \eqnum{1830} is in agreement
with \eqnum{880},
which holds when $D = 1$.

\subsubsection*{The helicoidal solution}

Compare the Gauss map \eqnum{980} of a CMC helicoid with the Purkait-Ray
solution \eqnum{350}
of the O(3) model.  The correspondence is obvious and we may relate the two
parameters
$\tw{D}$ and $h$ of the CMC helicoid to the constants $c_1$, $c_2$, $c_3$
appearing in
\eqnum{350}:
\beq{1840}
\tw{D}^2 = \frac{\beta^2}{\alpha^2}
	 = \frac{2c_1 + c_2^2 + 4c_3^2 + 2\delta}{2c_1 + c_2^2 + 4c_3^2 - 2\delta}
\eeq
where $\delta$ is given by \eqnum{370}, and
\beq{1850}
h = -4c_2^{-1}c_3.
\eeq
Therefore, when the pitch $h$ of the helicoid is non-zero, the Gauss map of a
helicoid of
constant mean curvature induces the Purkait-Ray solution of the O(3) sigma
model.  This is
the justification for calling \eqnum{350} the `helicoidal solution'.

\newpage
\section{Conclusion}
\label{conc}

In this paper we have shown that the Belavin-Polyakov soliton solutions of the
non-linear
O(3) sigma model in two dimensions are induced by the Gauss maps of multiple
coverings of the
unit sphere.  These surfaces have a total curvature $K_{\rm tot}$ equal to the
total energy
$E_{\rm tot} = 4\pi Q$, $Q > 0$, of the soliton configuration, and an Euler
characteristic
$\chi$ equal to twice the total topological charge $Q$.  We have shown that
antisoliton
solutions are the Gauss maps of certain minimal surfaces.  The total curvature
$K_{\rm tot}$
of these surfaces is the negative of the total energy: $K_{\rm tot} = -E_{\rm
tot} = 4\pi Q$,
$Q < 0$.

Other solutions, for which, in general, the total energy $E_{\rm tot}$ is
infinite, were
found to be the Gauss maps of surfaces of constant mean curvature.  The
differential equation
which is satisfied by the Gauss map of such a surface is the field equation of
the O(3)
model.	The total curvature $K_{\rm tot}$ is still equal to $4\pi Q$ but now the
topological
charge $Q$ need be neither integral nor finite.	 The elliptic interpolating
solution, in the
case of interpolation between merons and solitons, is induced by the Gauss map
of the
one-parameter family of Delaunay surfaces.  A more general solution, due to
Purkait and Ray,
was found to be the Gauss map of the two-parameter family of CMC helicoids
discovered by do
Carmo and Dajczer.

There exist other CMC surfaces in $R^3$, known as `Wente tori' \cite{wente86},
but there are
few explicit expressions for the position vector of such surfaces.  A number
are given in a
paper by Walter\cite{walter87} but their complicated algebraic form prohibits
easy use.  If
the Gauss map of a Wente torus could be deduced then it would correspond to a
new solution to
the O(3) model.  Also, it would be satisfying if a family of CMC surfaces could
be found
which corresponds to the interpolation between antimerons and antisolitons.

The obvious generalisation of the present work is to include time-dependence in
the sigma
model.  The field vector $\V{n}(x_1,x_2,t)$ could then be regarded in one of
two ways, either
as the normal to a moving two-dimensional surface in $\R^3$ or as the normal to
a three-%
dimensional hypersurface in $\R^4$.  Work in this direction could well prove
useful, for the following reason: time-dependence in the O(3) model causes new
topological
features to arise which could benefit from a surface-geometric interpretation.
For example,
the charge density $J(x_1,x_2)$ becomes one component of a topological current
$J_\mu(x_1,x_2,t)$.  Further, field configurations $\V{n}(x_1,x_2,t)$ which
tend to the same
constant value at {\it both} spatial and temporal infinity cause spacetime to
be
compactified to $\ess^3$.  The field is then a mapping from $\ess^3$ to
$\ess^2$, and such
mappings fall into homotopy classes according to
\[ \pi_3(\ess^2) \sim \Z. \]
The integer which labels these classes is the Hopf invariant\cite{flanders} $H$
which, in
the O(3) model, has the effect of imparting fractional spin and statistics to
the Belavin-%
Polyakov soliton after quantisation \cite{WZ83}.  But the literature on the
subject of the
Hopf invariant in the O(3) sigma model is rife with inaccuracies and seeming
inconsistencies,
although some papers are remedying this situation \cite{deazetal92,baezetal91}.

It may be that if the results presented here can be generalised to include
time-dependence
then a surface-geometric interpretation of quantities such as $J_\mu$ and $H$
can be found.
A similar idea involving the geometry of space curves has been used recently by
Balakrishnan
{\it et al}\cite{balaketal93b}.  These authors found that the current $J_\mu$
arises from anholonomy effects when a point $(x_1,x_2,t) \equiv (x,y,z)$ on a
three-parameter
space curve moves to $(x+\Delta x,y+\Delta y,z+\Delta z)$.  The field vector
$\V{n}$ is
identified with the tangent to the curve.  An analogous treatment in terms of
surfaces would
complement the work of Balakrishnan {\it et al.}\ and provide an alternative,
and perhaps
richer, geometric view of the structure of the time-dependent O(3) model.

\subsection*{Acknowledgements}

M. S. O. acknowledges financial support from the U. K. Science and Engineering
Research
Council under grant number 90806990.

\newpage
\section*{Appendix: The Gauss map of a CMC helicoid}

This appendix contains a derivation of equations \eqnum{980} which, when
combined with
\eqnum{840}, give the Gauss map $\zeta$ of a helicoid of constant mean
curvature.  If the
parameter $h$ is set to zero in the derivation then the Gauss map of a Delaunay
surface,
equations \eqnum{850}, is obtained.

We work initially in the $(s,\theta)$-parametrisation and later change to the
isothermal
coordinated $(\eta,\theta)$, where $\eta(s)$ is defined by \eqnum{970}.

{}From equations \eqnum{900}--\eqnum{940} it is seen firstly that the tangent
vectors to a
CMC helicoid are
\beq{1100}
\V{X}_s = \frac{1}{X Y}\bve Y D\cos s\cos\Delta + h(1 + D\sin s)\sin\Delta \\
			    Y D\cos s\sin\Delta - h(1 + D\sin s)\cos\Delta \\
			    X(1 + D\sin s) \eve \comma
\V{X}_\theta = \bve -X\sin\Delta \\ X\cos\Delta \\ h \eve
\eeq
and therefore that the unit normal vector is
\beq{1110}
\V{n} = \frac{-1}{X Y}\bve Y(1 + D\sin s)\cos\Delta - h D\cos s\sin\Delta \\
			   Y(1 + D\sin s)\sin\Delta + h D\cos s\cos\Delta \\
			   -X D\cos s \eve.
\eeq
To ensure consistency with the results obtained for spheres and minimal
surfaces we take the
form of the stereographic projection to be
\beq{1120}
\zeta = \frac{n_1 - \I n_2}{1 + n_3}
  = \E^{-\I(\Delta + \pi)} \left[\frac{Y(1 + D\sin s) - \I h D\cos s}{X(Y +
D\cos s)}\right].
\eeq
The numerator of the fraction in the bracket is converted to the
modulus-argument form
\beq{1130}
Y(1 + D\sin s) - \I h D\cos s \equiv \rho\E^{-\I\gamma}
\eeq
where
\bsub{1140}\bea
\rho^2 &\equiv& Y^2(1 + D\sin s)^2 + h^2 D^2\cos^2 s = X^2(Y^2 - D^2\cos^2 s)
\label{1140a}
		\\[2mm]
\gamma &\equiv& \tan^{-1}\left[\frac{h D\cos s}{Y(1 + D\sin s)}\right]
\nonumber \\[2mm]
       &=& -h D\int\frac{Y^2(D + \sin s) + D(1 + D\sin s)\cos^2 s}
			{X^2 Y(Y^2 - D^2\cos^2 s)}\,\dee s. \label{1140b}
\eea\esub
Then the Gauss map \eqnum{1120} becomes
\beq{1145}
\zeta = \E^{-\I\Phi}\left(\frac{1 - f}{1 + f}\right)^\half
\eeq
where
\bsub{1150}\bea
\Phi &=& \Delta + \pi + \gamma \nonumber \\
     &=& \theta + \pi - h\int\frac{\dee s}{Y(1 - f^2)} \label{1150a}
\eea
and
\beq{1150b}
f \equiv DY^{-1}\cos s.
\eeq\esub
Next, we change variables temporarily in order to simplify equation
\eqnum{1145} for $\zeta$.
The change of variables amounts essentially to inverting \eqnum{970}, \ie to
finding $s$ in
terms of $\eta$.  The integral in \eqnum{970} is elliptic and may be written in
terms of the
standard elliptic integral
\beq{1160}
t(s) \equiv \int_0^{\phi(s)}\frac{\dee x}{(1 - k^2\sin^2 x)^\half}.
\eeq
In fact we have \cite{gradryzh}
\beq{1170}
\eta = -2A^{-1}t
\eeq
\beq{1180}
\phi(s) \equiv \sin^{-1}\left(\frac{1 - \sin s}{2}\right)^\half
\eeq
\beq{1190}
k^2 \equiv \frac{4\tw{D}}{(1 + \tw{D})^2}
\eeq
where $A$ and $\tw{D}$ are defined by equations \eqnum{990}.  Now, by
definition of the
Jacobi elliptic functions, we have
\beq{1200}
\sn(t|k) = \sin\phi
\eeq
and therefore \eqnum{1180} gives
\bsub{1210}\bea
\sin s &=& 1 - 2\,\sn^2(t|k) \label{1210a} \\
\cos s &=& 2\,\sn(t|k)\,\cn(t|k). \label{1210b}
\eea\esub
Since, by \eqnum{1170}, $t$ is a function of $\eta$, either of these relations
\eqnum{1210}
may be taken as defining the function $s(\eta)$ which is the inverse of
$\eta(s)$.  We now
change temporarily to the variables $(t,\theta)$.  It follows from
\eqnum{1190}, \eqnum{990},
and the relation
\beq{1220}
\dn^2(t|k) = 1 - k^2\sn^2(t|k)
\eeq
that
\beq{1230}
Y = A\,\dn(t|k).
\eeq
Therefore equations \eqnum{1210} yield
\beq{1240}
\dee s = -2\,\dn(t|k)\,\dee t = -2A^{-1}Y\,\dee t.
\eeq
Consequently, equations \eqnum{1150} for the quantities $\Phi$ and $f$ become
\bsub{1250}
\beq{1250a}
\Phi = \theta + \pi + \frac{2h}{A}\int\frac{\dee t}{1 - f^2}
\eeq
\beq{1250b}
f = \left(\frac{2D}{A}\right)\frac{\sn(t|k)\,\cn(t|k)}{\dn(t|k)}.
\eeq\esub
We now invoke the `ascending Landen transformation' for the Jacobi function
$\sn$\,:
\beq{1260}
\sn(x|m) = \left(\frac{2}{1+m}\right)\frac{\sn(y|n)\,\cn(y|n)}{\dn(y|n)}
\eeq
where
\bsub{1270}
\beq{1270a}
y \equiv \left(\frac{1+m}{2}\right)x
\eeq
\beq{1270b}
n^2 \equiv \frac{4m}{(1 + m)^2}.
\eeq\esub
If the identifications
\beq{1280}
y = t \comma n = k \comma m = \tw{D}
\eeq
are made then equation \eqnum{1270b} coincides with \eqnum{1190} and
\eqnum{1260} becomes, on
using \eqnum{990},
\beq{1290}
\sn(x|\tw{D}) = \left(\frac{A+B}{A}\right)\frac{\sn(t|k)\,\cn(t|k)}{\dn(t|k)}.
\eeq
We use \eqnum{1290} and \eqnum{990} to bring \eqnum{1250b} to the form
\beq{1300}
f = \beta\,\sn(x|\tw{D}).
\eeq
Now we revert to the variables $(\eta,\theta)$ by finding $x$ in terms of
$\eta$ from
\eqnum{1270a} and \eqnum{990}:
\beq{1310}
x = \left(\frac{2}{1+\tw{D}}\right)t = -\alpha\eta.
\eeq
Therefore the quantity $f$ becomes
\beq{1320}
f = -\beta\,\sn(\alpha\eta|\tw{D}).
\eeq
Finally, we bring together equations \eqnum{1145}, \eqnum{1250a} and
\eqnum{1320} to find
that the Gauss map $\zeta$ of a CMC helicoid takes the form \eqnum{840} with
$R$ and $\Phi$
given by \eqnum{980}, as claimed.


\begin{thebibliography}{10}

\bibitem{eisenhart}
L.~P. Eisenhart,
\newblock {\em A Treatise on the Differential Geometry of Curves and Surfaces},
\newblock Dover, New York, 1960.

\bibitem{gardneretal67}
C.~S. Gardner, J.~M. Greene, M.~D. Kruskal, and R.~M. Miura,
\newblock {\em Phys.\ Rev.\ Lett.} {\bf 19}, 1095--1097 (1967).

\bibitem{lund78}
F.~Lund,
\newblock {\em Ann.\ Phys.} {\bf 115}, 251--268 (1978).

\bibitem{lundregge76}
F.~Lund and T.~Regge,
\newblock {\em Phys.\ Rev.\ D} {\bf 14}, 1524--1535 (1976).

\bibitem{ablowitzetal74}
M.~J. Ablowitz, D.~J. Kaup, A.~C. Newell, and H.~Segur,
\newblock {\em Stud.\ Appl.\ Math.} {\bf 53}, 249--315 (1974).

\bibitem{gursesnutku81}
M.~G\"{u}rses and Y.~Nutku,
\newblock {\em J. Math.\ Phys.} {\bf 22}, 1393--1398 (1981).

\bibitem{sym82}
A.~Sym,
\newblock {\em Lett.\ Nuovo Cimento} {\bf 33}, 394--400 (1982).

\bibitem{sym85}
A.~Sym,
\newblock {\em Lecture Notes in Physics} {\bf 239}, 154--231 (1985).

\bibitem{balachandranetal}
A.~P. Balachandran, G.~Marmo, B.~S. Skagerstam, and A.~Stern,
\newblock {\em Classical Topology and Quantum States},
\newblock World Scientific, Singapore, 1991.

\bibitem{zak}
W.~J. Zakrzewski,
\newblock {\em Low-Dimensional Sigma Models},
\newblock Adam Hilger, Bristol, 1989.

\bibitem{lund77}
F.~Lund,
\newblock {\em Phys.\ Rev.\ D} {\bf 15}, 1540--1543 (1977).

\bibitem{houetal85}
B.~Y. Hou, B.~Y. Hou, and P.~Wang,
\newblock {\em J. Phys.\ A} {\bf 18}, 165--185 (1985).

\bibitem{ody93}
M.~S. Ody,
\newblock {\em The (2+1)-Dimensional Non-Linear O(3) Sigma Model and the
  Classical Differential Geometry of Curves and Surfaces},
\newblock Ph.\ {D}. thesis, University of Kent at Canterbury, 1993.

\bibitem{rajaraman}
R.~Rajaraman,
\newblock {\em Solitons and Instantons},
\newblock North-Holland, Amsterdam, 1982.

\bibitem{flanders}
H.~Flanders,
\newblock {\em Differential Forms with Applications to the Physical Sciences},
\newblock Dover, New York, 1989.

\bibitem{belapoly75}
A.~A. Belavin and A.~M. Polyakov,
\newblock {\em JETP Lett.} {\bf 22}, 245--247 (1975).

\bibitem{garberetal79}
W.~D. Garber, S.~N.~M. Ruijsenaars, E.~Seiler, and D.~Burns,
\newblock {\em Ann.\ Phys.} {\bf 119}, 305--325 (1979).

\bibitem{gross78}
D.~J. Gross,
\newblock {\em Nucl.\ Phys.\ B} {\bf 132}, 439--456 (1978).

\bibitem{dealfaroetal78}
V.~de~Alfaro, S.~Fubini, and G.~Furlan,
\newblock {\em Nuovo Cimento A} {\bf 48}, 485--499 (1978).

\bibitem{callanetal77}
C.~G. Callan, R.~F. Dashen, and D.~J. Gross,
\newblock {\em Phys.\ Lett.\ B} {\bf 66}, 375--381 (1977).

\bibitem{takhom81}
S.~Takeno and S.~Homma,
\newblock {\em Prog.\ Theor.\ Phys.} {\bf 65}, 1844--1857 (1981).

\bibitem{tseitlin83}
M.~G. Tseitlin,
\newblock {\em Theor.\ Math.\ Phys.} {\bf 57}, 1110--1117 (1983).

\bibitem{abrasteg}
M.~Abramowitz and I.~A. Stegun,
\newblock {\em Handbook of Mathematical Functions},
\newblock Dover, New York, 1970.

\bibitem{khare80}
A.~Khare,
\newblock {\em J. Phys.\ A} {\bf 13}, 2253--2259 (1980).

\bibitem{ghikavis82}
G.~Ghika and M.~Visinescu,
\newblock {\em Z. Phys.\ C} {\bf 11}, 353--357 (1982).

\bibitem{abbott82}
R.~B. Abbott,
\newblock {\em Z. Phys.\ C} {\bf 15}, 51--59 (1982).

\bibitem{purkaitray86}
S.~Purkait and D.~Ray,
\newblock {\em Phys.\ Lett.\ A} {\bf 116}, 247--250 (1986).

\bibitem{struik}
D.~J. Struik,
\newblock {\em Lectures on Classical Differential Geometry},
\newblock Dover, New York, second edition, 1988.

\bibitem{hopf83}
H.~Hopf,
\newblock {\em Lecture Notes in Mathematics} {\bf 1000}, 77--184 (1983).

\bibitem{nitsche}
J.~C.~C. Nitsche,
\newblock {\em Lectures on Minimal Surfaces}, volume~1,
\newblock Cambridge University Press, Cambridge, 1989.

\bibitem{hoffoss85}
D.~A. Hoffman and R.~Osserman,
\newblock {\em Proc.\ London Math.\ Soc.} {\bf 50}, 27--56 (1985).

\bibitem{ruhvilms70}
E.~A. Ruh and J.~Vilms,
\newblock {\em Trans.\ Am.\ Math.\ Soc.} {\bf 149}, 569--573 (1970).

\bibitem{kenmotsu79}
K.~Kenmotsu,
\newblock {\em Math.\ Ann.} {\bf 245}, 89--99 (1979).

\bibitem{eells87}
J.~Eells,
\newblock {\em Math.\ Intelligencer} {\bf 9}, 53--57 (1987).

\bibitem{kenmotsu80}
K.~Kenmotsu,
\newblock {\em T\^{o}hoku Math.\ J.} {\bf 32}, 147--153 (1980).

\bibitem{docdaj82}
M.~P. do~Carmo and M.~Dajczer,
\newblock {\em T\^{o}hoku Math.\ J.} {\bf 34}, 425--435 (1982).

\bibitem{wente86}
H.~C. Wente,
\newblock {\em Pacific J. Math.} {\bf 121}, 193--243 (1986).

\bibitem{walter87}
R.~Walter,
\newblock {\em Geometriae Dedicata} {\bf 23}, 187--213 (1987).

\bibitem{WZ83}
F.~Wilczek and A.~Zee,
\newblock {\em Phys.\ Rev.\ Lett.} {\bf 51}, 2250--2252 (1983).

\bibitem{deazetal92}
J.~A. de~Azc\'{a}rraga, J.~M. Izquierdo, and W.~J. Zakrzewski,
\newblock {\em J. Math.\ Phys.} {\bf 33}, 1272--1280 (1992).

\bibitem{baezetal91}
J.~C. Baez, A.~R. Bishop, and R.~Dandoloff,
\newblock {\em Mod.\ Phys.\ Lett.\ B} {\bf 5}, 2003--2005 (1991).

\bibitem{balaketal93b}
R.~Balakrishnan, A.~R. Bishop, and R.~Dandoloff,
\newblock {\em Phys.\ Rev.\ B} {\bf 47}, 5438--5441 (1993b).

\bibitem{gradryzh}
I.~S. Gradshteyn and I.~M. Ryzhik,
\newblock {\em Table of Integrals, Series, and Products},
\newblock Academic Press, New York, 1980.

\end{thebibliography}
\end{document}